\documentclass[11pt,a4paper]{article}
\usepackage{amssymb}
\usepackage[fleqn]{amsmath}
\usepackage{graphicx}
\usepackage{epsfig}

\usepackage{amsmath}
\usepackage{amsfonts}
\usepackage{amssymb}
\begin{document}
\title{\textsc{
Dynamics of Social Systems: Cooperation and Free-Riding }}
\author {Yiping Ma (1, 2) \and Mirta B. Gordon (1) \and Jean-Pierre Nadal (3)\\}
\vspace{1 cm}
\date{ \scriptsize{
(1) Laboratoire Leibniz-IMAG, Grenoble\\
46, Ave. F\'elix Viallet, 38031 Grenoble Cedex 1, France\\
(mirta.gordon@imag.fr, web: http//www-leibniz.imag.fr/Apprentissage)\\
\bigskip
(2) Department of Physics, Hong Kong University of Science and
Technology \\
Clear Water Bay, Kowloon, Hong Kong \\
(ph\_myp@stu.ust.hk, http://ihome.ust.hk/\~{}ph\_myp/) \\
\bigskip
(3) Laboratoire de Physique Statistique, Ecole Normale Sup\'erieure, \\
24 rue Lhomond, 75231 Paris cedex 05, France\\
(nadal@lps.ens.fr, http//www.lps.ens.fr/\~{ }nadal)\\}
}
\maketitle \thispagestyle{empty} \vspace{-1 cm}
\bigskip
\centerline  {\date{\today} }
\begin{abstract}
\scriptsize{We study the mean field dynamics of a model introduced
by Phan et al [Wehia, 2005] of a polymorphic social community. The
individuals may choose between three strategies: either not to
join the community or, in the case of joining it, to cooperate or
to behave as a free-rider. Individuals' preferences have an
idiosyncratic component and a social component. Cooperators bear a
fixed cost whereas free-riders support a cost proportional to the
number of cooperators. We study the dynamics of this model
analytically in the mean field approximation for both parallel and
sequential updating. As we vary one of the parameters while
keeping the other parameters fixed, the phase diagram experiences
a rich class of bifurcations. Noticeably, a limit cycle is shown
to exist in both parallel and sequential updating, under certain
parameter settings. A comparison of the analytical
predictions with computer simulations is also included. \\

 }

\end{abstract}

\newpage

\section{Introduction}
Recently \cite{PhWaGoNa05,GoPhWaNa05} a model was proposed to
analyze social organizations whose members are expected to
cooperate to a public good. Basic evidence on several kinds of
communities, as well as data obtained in public goods experiments
\cite{FehrGachter2002,GaechterFehr1999}, reveal a rough partition
between individuals that cooperate to the public good and pure
consumers (also called free-riders). This polymorphic
configuration seems to be a stable form of organization.

The model corresponds to the following situation: individuals
have to decide whether to join an organization or community. The members of
the organization have to contribute to a task whose realization is
beneficial for everybody. Cooperators bear a fixed cost for
producing the public good for the community. The surplus of all
the individuals, cooperators or not, increases proportionally to the number of
cooperators. Individuals that do not cooperate are punished by
cooperators through costless moral disapproval, which may be
either a subjective moral burden, or a true sanction. This cost,
proportional to the fraction of cooperators, is idiosyncratically
weighted.

In the present paper, we consider the dynamical evolution of the
model analytically, within a mean field approximation. Parallel
and sequential updating are studied through the evolution of the
corresponding map and flow respectively. They present different
behaviors, depending on the models parameters.

The most striking result of our analysis is that for some range of
parameter values and initial conditions, the trajectories of the
flow reach the fixed points determined by Phan et al.
\cite{PhWaGoNa05} through quasi-cyclic paths that may last for
very long times. The corresponding map exhibits a limit cycle.
Thus, under such conditions, macroscopic fractions of individuals
change their strategies successively. In experimental economics
settings that correspond to parallel updating, true cycles might
appear. In sequential updating, the transient oscillatory behavior
of the system may be very long lasting, and should thus be
observable in actual systems.

The paper is organized as follows: next section presents the
details of the model and its equilibrium fixed points.
Section \ref{sec:mean fiels dynamics} is devoted to the study of
a mean field approximation of the dynamical equations. These describe
the evolution of the fraction of cooperators and of free-riders.
We consider both parallel and sequential updating, and study the
corresponding map and flow. We show that the system may exhibit
strong oscillations of the fraction of cooperators and of free-riders,
and that limit cycles may exist for some range of the parameters.
Numerical simulations presented in section \ref{sec:computer simulations}
show that the predicted behaviours exist in finite size systems and
might be observable in actual systems. The paper ends with a discussion
and some conclusions.

\section{The model}
\label{sec:the model}
The basic economic model analyzed in this paper was introduced by
Phan et al. \cite{PhWaGoNa05}. It considers a system of $N$ agents
that must choose one among the following strategies:
\begin{eqnarray}
S_i &=& 1 \;\;\;\; \text{(to join the community and cooperate)} \nonumber\\
S_i &=& 2 \;\;\;\; \text{(to join the community and free-ride)} \nonumber \\
S_i &=& 3 \;\;\;\; \text{(not to join the community)} \nonumber
\end{eqnarray}
Each individual $i$ has a private idiosyncratic willingness to
join the community, $h_i$. Its mean value over the population is
$h$ and we write: $h_i \equiv h + y_i$ where $y_i$ is a quenched
random variable with zero mean. Individuals that join the
community have a social benefit proportional to the fraction of
the population that join the community, weighted by a coefficient
$j$, and an additional payoff proportional to the fraction of
cooperators, weighted by a constant $g$. Then, depending on
whether they cooperate or not, they bear different costs. A
cooperator bears a fixed cost $c$. Free-riders do not bear this
cost, but instead support a moral punishment proportional to the
fraction of cooperators. This punishment is weighted by an
idiosyncratic positive constant $x_i$. All the parameters of the
model ($h,j,g,c$ as well as the values of $\{y_i\}$ and $\{x_i\}$
for $1 \leq i \leq N$) are measured in units of the variance of
the random variable $y_i$.

The fraction of cooperators is:
$$\eta_c \equiv \frac{1}{N} \sum_{i=1}^N \delta_{S_i,1},$$
where $\delta$ denotes the Kroenecker delta. That of free-riders is:
$$\eta_f \equiv \frac{1}{N} \sum_{i=1}^N \delta_{S_i,2},$$
so that the fraction of individuals that join the community is $\eta_a=\eta_c+\eta_f \leq 1$.

Each agent chooses the strategy that maximizes his surplus function,
\begin{equation}\label{eq.surplus}
V_i \left( {S_i } \right) \; = \; ( A_i + B_i ) \; \delta_{S_i,1}
\;+\; A_i  \; \delta_{S_i,2}.
\end{equation}
$A_i$ is the surplus of joining the community being free-rider, and $B_i$ is the bonus
of cooperating:
\begin{subequations}\label{eq.utilities}
  \begin{gather}
  A_i = h + y_i +j (\eta_c+\eta_f) + (g -x_i) \eta_c, \label{eq.U_adopt}   \\
  B_i = x_i \eta_c - c.                               \label{eq.U_coop}
  \end{gather}
\end{subequations}
From equation(\ref{eq.surplus}), the best response of agent $i$ to
the neighborhood's behavior can be rewritten as follows:
\begin{subequations}
\label{eq.strategiesEquilibrium}
  \begin{gather}
    S_i= 1, \; \Longleftrightarrow A_i+B_i>0 \; {\rm and} \; B_i>0; \label{eq.11} \\
    S_i= 2, \; \Longleftrightarrow A_i>0 \;     {\rm and} \; B_i<0; \label{eq.10} \\
    S_i= 3, \; {\rm otherwise} \label{eq.0}
  \end{gather}
\end{subequations}

\subsection{The Cluster Structure}
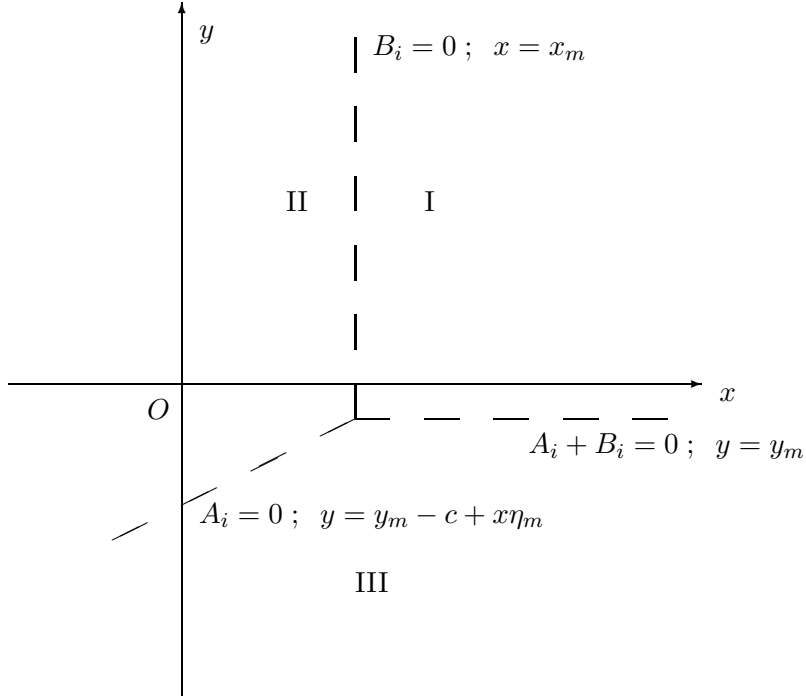
\begin{figure}[!hbp]
\setlength{\unitlength}{0.016\textwidth}
\begin{center}
\begin{picture}(40,40)(-18,-18)
    \put(-20,0){\vector(1,0){40}}
    \put(21,-1){$x$}
    \put(-10,-18){\vector(0,1){40}}
    \put(-9,20){$y$}
    \put(-12,-2){$O$}
    \multiput(0,-2)(4,0){5}{\line(1,0){2}}
    \put(10,-4){$A_i+B_i=0 \; ; \;\; y=y_m$}
    \multiput(0,-2)(0,4){6}{\line(0,1){2}}
    \put(1,19){$B_i=0 \; ; \;\; x=x_m$}
    \multiput(0,-2)(-4,-2){4}{\line(-2,-1){2}}
    \put(-9,-8){$A_i=0 \; ; \;\; y=y_m - c + x \eta_m$}
    \put(4,10){I}
    \put(-4,10){II}
    \put(0,-12){III}
\end{picture}
\end{center}
\caption{Boundaries between regions corresponding to the different
individual choices, as a function of the values of the quenched
random variables.} \label{fig.cluster_choice}
\end{figure}

In order to get some insight about the problem, it is useful to
consider a plane whose axes are the quenched random variables $x$
(abscissas) and $y$ (ordinates), as in figure
\ref{fig.cluster_choice}.  Each individual $i$ is represented by a
point according to his values $(x_i,y_i)$. The lines represent the
marginal individuals whose utilities are at the boundaries between
different optimal strategies. These are:
\begin{eqnarray}
\label{eq.marginals}
y & = &  y_{m} \equiv -h-j (\eta_c+\eta_f) - g \eta_c + c,  \\
y & = &  y_m-c+ x \eta_c,  \\
x & = &  x_m \equiv \frac{c}{\eta_c}.
\end{eqnarray}
Then, according to (\ref{eq.strategiesEquilibrium}), individuals in region I will choose to
cooperate, in region II to free-ride and in region III not to join the community.

\subsection{Fraction of cooperators and free-riders at equilibrium}
The fixed points of the system for different distributions of the
quenched random variables $\{x_i,y_i\}$ have been discussed in
\cite{PhWaGoNa05} and \cite{GoPhWaNa05} in the mean field
approximation $N \rightarrow \infty$. We summarize them here for
completeness.

Introducing the complementary cumulative functions:
\begin{equation}
G_\chi (\zeta ) \equiv 1-F_\chi (\zeta ) = \int_\zeta ^{\infty} f_\chi (\xi ) d\xi
\label{eq.G}
\end{equation}
where $\chi \in \{X,Y\}$, and $F_\chi(\zeta)$ is the cumulative
distribution of $f_\chi$, $\eta_c$ and $\eta_f$ have the following
expressions in terms of the $G$ functions and the marginal values
(\ref{eq.marginals}):
\begin{subequations}
\label{eq.etas}
  \begin{gather}
    \eta_c = G_X(x_m) \; G_Y(y_m), \label{eq.etac} \\
    \eta_f = \int_{0}^{x_m} f_X(x) \; G_Y(y_m-c+x \eta_c)\; dx, \label{eq.etaf}.
  \end{gather}
\end{subequations}
The solutions to equations (\ref{eq.etas}) give the fraction of
cooperators and free-riders at equilibrium as a function of the
parameters of the problem, namely $h, j, g, c, d$ (these
parameters are measured in units of the variance of the
distribution of $y$). Notice that $\eta_c=0$ is always solution of
(\ref{eq.etac}), in which case all the members are free-riders,
and the problem reduces to that of the simple social interactions
model considered in \cite{Durlauf01,GoNaPhVa05} and references
therein. It is important to stress that, due to the non symmetric
interaction $x_i$, there is no reason that the stationary states
of the system be exclusively fixed points, as is the case for
symmetric interactions\footnote{In Ising systems, with binary
microscopic states and symmetric interactions, the only attractors
are fixed points for sequential dynamics, and either fixed points
or cycles of order two for parallel dynamics.}.

In this paper we consider the most interesting of the cases
analyzed in \cite{PhWaGoNa05}, with the $x_i$ following a uniform
distribution of finite width $d$:
\begin{subequations}
\label{eq.f_X}
  \begin{gather}
  f_X(x)=\frac{1}{d} \;\;\;  {\rm for} \;\;\;  0 \leq x \leq d, \\
  f_X(x)=0  \;\;\; {\rm otherwise}.
  \end{gather}
\end{subequations}
and the $y_i$ distributed according to the following probability density function:
\begin{equation}
f_Y(y)=\frac{1}{4 \cosh^2 (y/2)}. \label{eq.f_Y}
\end{equation}
The cumulative function corresponding to (\ref{eq.f_Y}) is the logistic distribution, and its complementary function is:
\begin{equation}
G_Y(z) = 1/[ 1 + \exp(z) ].
\label{eq.logit}
\end{equation}
In this case, following Phan et al. \cite{PhWaGoNa05}, equations (\ref{eq.etas}) may be written as follows:
\begin{subequations} \label{eq.eta_unif}
  \begin{gather}
\eta_c = ( 1 - \frac{\rho}{\eta_c }) \; G_Y(c - Z) \label{eq.eta_c_unif} \\
\eta_f = \frac{\rho}{c \eta_c } \; \log [1 + (e^c - 1) G_Y(c-Z)], \label{eq.eta_f_unif}
\end{gather}
\end{subequations}
where $\rho \equiv c/d$ and $Z \equiv h + j (\eta_c+\eta_f)+ g \eta_c$.

Calling $G$ the value taken by $G_Y(c - Z)$, one has $\eta_c = ( 1
- \frac{\rho}{\eta_c }) \; G$, which gives $\eta_c$ (if $>0$) in
terms of $G$:
\begin{equation}
\eta_c = \eta_c^{\pm }[G] \equiv \frac{1}{2} G \; \{ 1 \pm   [1 - \frac{4\rho }{G }]^{1/2} \}
\label{eq_etac_pm}
\end{equation}
Later we'll see that both the $+$ and the $-$ branches may give
stable equilibria, though the parameter range for which a stable
equilibrium exists for the $-$ branch is much narrower than for
the $+$ branch. The function $\eta_c^+[G]$ is very close to its
asymptote, $G - \rho$, for all the values of $G$. From
(\ref{eq_etac_pm}) one gets that if $\eta_c \neq 0$, then $\eta_c
\geq 2 \rho$. Equality can occur when $j=g=0$.

\begin{figure}[!hbp]
\centering
\includegraphics[width=0.6\textwidth]{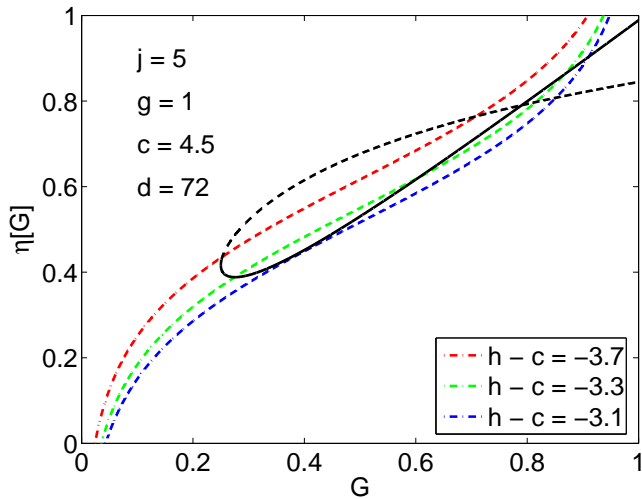}
\caption[]{Solution by curve intersection - a case of interest.}
\label{fig.GraphicSolution}
\end{figure}

Equation (\ref{eq.eta_f_unif}) can also be parameterized in terms of $G$ by defining
\begin{equation}
\eta \equiv  \frac{j  (\eta_c+\eta_f)\;+\; g \eta_c }{j\;+\; g}.
\label{eq.eta}
\end{equation}
Introducing (\ref{eq_etac_pm}) and (\ref{eq.eta_f_unif}) into (\ref{eq.eta}) we obtain an equation for $\eta$:
\begin{equation}
\eta = \eta_1[G] \equiv \; \eta_c^{\pm }[G] + \frac{j}{j+g}\; \frac{1}{\eta_c^{\pm }[G]}\;\frac{1}{d} \log [1 + (e^c - 1) G]
\label{eq.etat1}
\end{equation}
Inverting (\ref{eq.logit}) for $z= c - h - (j+g) \eta$, we obtain $\eta$ in terms of $G$:
\begin{equation}
\label{eq.etat2}
\eta = \eta_2[G] \equiv \;  \frac{1}{j+g} \; \{ c - h - \log\frac{1-G}{G} \}.
\end{equation}
The possible solutions $\eta$ are then obtained by the intersects
of the curves $\eta_1[G]$ and $\eta_2[G]$. Introducing the
corresponding value of $G$ into (\ref{eq_etac_pm}) allows to
determine $\eta_c$, and then $\eta_f$ is deduced by introducing
the values of $\eta_c$ and $\eta$ into (\ref{eq.eta}). An example
of curves $\eta_1[G]$ and $\eta_2[G]$ ($G \in [0,1]$) is shown on
Figure (\ref{fig.GraphicSolution}), which presents up to 3
intersects (for $h-c=-3.3$), although one can check that at most 2
correspond to stable equilibria. Notice that this analysis allows
to determine the fixed points of the system, but doesn't give any
hint about the existence of cycles.

\section{Mean field dynamics}
\label{sec:mean fiels dynamics}
We are interested in the temporal evolution of the system within a
repeated game setting, in which individuals have to choose their
best strategies $S_i(t)$ based on the available information. We
assume that each time an agent has to make a decision, he has the
exact information of the global proportions of cooperators and
free-riders at the preceding outcome, $\eta_c(t-1)$ and
$\eta_f(t-1)$, and uses these quantities to estimate his utility
(\ref{eq.utilities}). Such a dynamics is called Cournot best reply
in economics literature. In simulations, starting from initial
guesses $\eta_c(0)$ and $\eta_f(0)$, the updating is said to be in
parallel if all the individuals in the population first determine
their best strategies based on the preceding outcome, and make
their decisions simultaneously afterwards. At the opposite, in
random sequential updating, a single individual is selected at
random and asked to make his decision at each time step. The
latter dynamics simulates systems where the individuals make their
decisions without any time coherence. In order to compare the time
scales of both dynamics, it is usual to consider that $N$
sequential time steps are equivalent to one parallel update.
Intermediate updating schemes may be implemented, but we only
consider these two extreme cases in this paper.

Referring back to Figure \ref{fig.cluster_choice}, since the
boundary lines depend on the values of $\eta_c(t)$ and
$\eta_f(t)$, they will shift in the course of updating according
to the perceived proportions of cooperators and free-riders.
Individuals whose values of $y_i$ and $x_i$ lie close to the
boundaries are susceptible to small changes of $\eta_c$ and
$\eta_f$: their strategies may change with time, and in turn
induce changes in those of the others.

We are interested in the way the system reaches its stable states
upon successive updates, that is, the path followed by a point
representative of the system's state in the plane $(\eta_c,
\eta_f)$. Such paths should either end up in one of the stable
fixed points determined by the previous analysis, or get trapped
in other types of attractors if they exist. In the following we
consider separately the two updating schemes, since the
corresponding dynamic equations are different.

\subsection{Parallel dynamics: two-dimensional map}
\label{sec:map}
In parallel dynamics, one assumes that the agents are updated
simultaneously starting from an arbitrary initial configuration,
i.e. the perceived $\eta_c$ and $\eta_f$ used by all the agents to
estimate their utilities are the same. The boundary lines between
the three regions of figure \ref{fig.cluster_choice}, which depend
on $\eta_c(t)$ and $\eta_f(t)$, partition the population at time
$t$ according to their estimated utilities: all the individuals
whose idiosyncratic parameters lie in region I will choose to
cooperate, those in region II to free-ride and those in region III
not to join. In this way, using equations (\ref{eq.etas}), the
dynamics can be formulated as the following two-dimensional
deterministic map in the simplex
$\mathbf{S}:=\{(\eta_c,\eta_f):\eta_c \ge 0,\eta_f \ge
0,\eta_c+\eta_f \leq 1\}$:
\begin{subequations}\label{eq:the_map}
\begin{gather}
    \Omega:(\eta_c',\eta_f')=(p(\eta_c,\eta_f),q(\eta_c,\eta_f))\\
    \qquad \equiv(\int_{x_m}^\infty f_X(x)dx\int_{y_m}^\infty
    f_Y(y)dy,\int_{-\infty}^{x_m}\int_{y_m-c+x\eta_c}^\infty
    f_X(x)f_Y(y)dydx).
\end{gather}
\end{subequations}
In the case considered in this paper, the density functions are
(\ref{eq.f_X}) and (\ref{eq.f_Y}). Since $f_X$ has a bounded
support, the first equation vanishes if $0<\eta_c(t)< \rho$. Thus,
if the initial value of $\eta_c$ is such that $\eta_c(0)< \rho$,
then $\eta_c(1)=0$ and a state with no cooperator is reached after
a single parallel update. Afterwards, on the axis $\eta_c=0$,
$\eta_f$ evolves according to
\begin{equation}\label{eq:eta_f_evolve}
    \eta_f(t+1)=\int_{-h-j\eta_f(t)}^\infty f_Y(y)dy.
\end{equation}
The equilibrium value of $\eta_f$ is obtained replacing
$\eta_f(t)$ by $\eta_f$ in the above equation. It has been shown
\cite{NaPhGoVa05} that there is a critical value of $j$, $j_B$,
such that for $j<j_B$, there is a single equilibrium. In that case
all the points $\eta_c(0) < \rho$ will eventually be mapped to it.
If $j>j_B$ equation (\ref{eq:eta_f_evolve}) with
$\eta_f(t)=\eta_f$ has three solutions. One of them, $\eta_{fu}$
is an unstable fixed point separating the basin of attraction of
the two others, that are stable. To summarize, when $\eta_c(0) <
\rho$, after a first time step the system is mapped to the axis
$\eta_c=0$ and then evolves, following (\ref{eq:eta_f_evolve}),
either to one fixed point (if $j<j_B$) or to one of two fixed
points (if $j>j_B$) depending on whether $\eta_f(0) > \eta_{fu}$
or $\eta_f(0) < \eta_{fu}$. In the case of the logistic
distribution considered here, $j_B=4$.

When $\eta_c(0) > \rho$, the dynamics is most
fruitfully studied numerically. In two-dimensional maps or flows,
a fixed point is called a sink if all the points in its
neighborhood converge to it, a source if these points diverge
from it and a saddle if in one direction the map or flow converges,
while in the other directions it diverges. Together these three types are called
hyperbolic, which is the only kind of
equilibria possibly encountered in a structurally stable system,
i.e. a system stable with respect to small variations of the
parameters (in our case $h$, $j$, $g$, $c$ and $d$). For our two-dimensional
map, the nature of an equilibrium point $(\eta_{c0},\eta_{f0})$ is determined
by the following Jacobian matrix
\begin{displaymath}\label{eq:jac_mat}
    \mathbf{J}_m=\left[\begin{array}{cc}
    \partial p(\eta_{c0},\eta_{f0})/\partial\eta_c &
    \partial p(\eta_{c0},\eta_{f0})/\partial\eta_f\\
    \partial q(\eta_{c0},\eta_{f0})/\partial\eta_c &
    \partial q(\eta_{c0},\eta_{f0})/\partial\eta_f
    \end{array} \right].
\end{displaymath}
Denoting the two eigenvalues of $\mathbf{J}_m$ as $a_1$ and $a_2$,
then the fixed point is a sink if $|a_1|<1$ and $|a_2|<1$, a
source if $|a_1|>1$ and $|a_2|>1$ and a saddle if
$(|a_1|-1)(|a_2|-1)<0$ (\cite{GuckenheimerHolmes1990}\S1.4). There
are two types of curves emanating from each saddle point $x_0$,
namely the stable manifold, defined as
\begin{equation}\label{eq:def_stab_mani}
    W_s(x_0)=\{x\in\mathbb{R}^2|\Omega^n(x)\rightarrow x_0 \textrm{ as
    } n\rightarrow\infty\}
\end{equation}
and the unstable manifold, defined as
\begin{equation}\label{eq:def_unst_mani}
    W_u(x_0)=\{x\in\mathbb{R}^2|\Omega^{-n}(x)\rightarrow x_0 \textrm{ as
    } n\rightarrow\infty\}
\end{equation}
$\Omega^{-1}$ denoting a backward iterate of the map. The
boundaries of the basins of attraction are usually formed by
stable manifolds. Moreover, if a transversal intersection between
the stable and unstable manifolds of the same saddle point exists,
the map will exhibit chaotic behavior.

\begin{figure}[!hbp]
\centering
\includegraphics[width=0.6\textwidth]{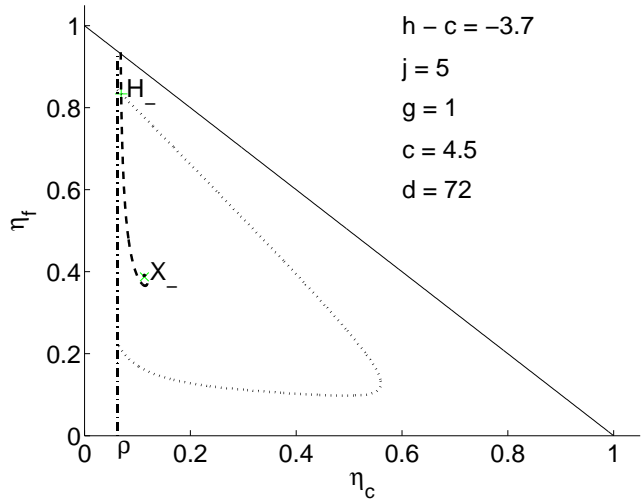}
\caption[]{Map. The stable and unstable manifolds for $h-c=-3.7$.
Throughout this article, stable manifolds are represented by
dashed lines and unstable manifolds by dotted lines. It's well
understood from their definitions that an arbitrary initial point
on the stable manifold will be mapped towards the saddle from
which it emanates, while one on the unstable manifold will be
mapped away from it. Sinks, sources and saddles are represented by
O, X, and H respectively, and the subscripts describe the branch
($\eta_1^+(G)$/$\eta_1^-(G)$) on which the equilibrium is found.}
\label{fig.map.-3.7}
\end{figure}

\begin{figure}[!hbp]
\centering
\includegraphics[width=0.6\textwidth]{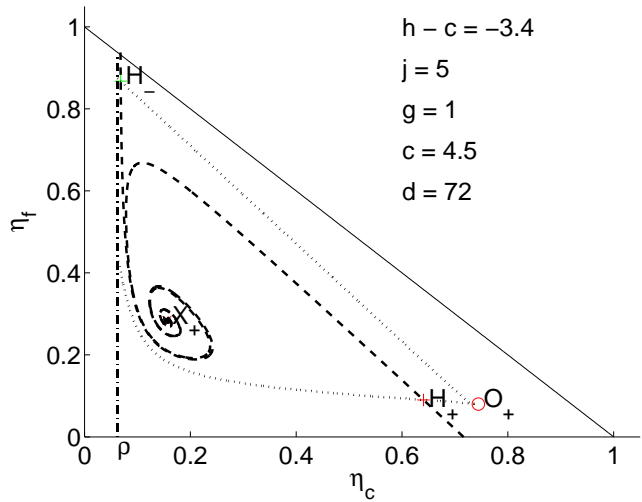}
\caption[]{Map. $H_+$ and $O_+$ have been created from a saddle node
bifurcation on $\eta_1^+(G)$.} \label{fig.map.-3.4}
\end{figure}

\begin{figure}[!hbp]
\centering
\includegraphics[width=0.6\textwidth]{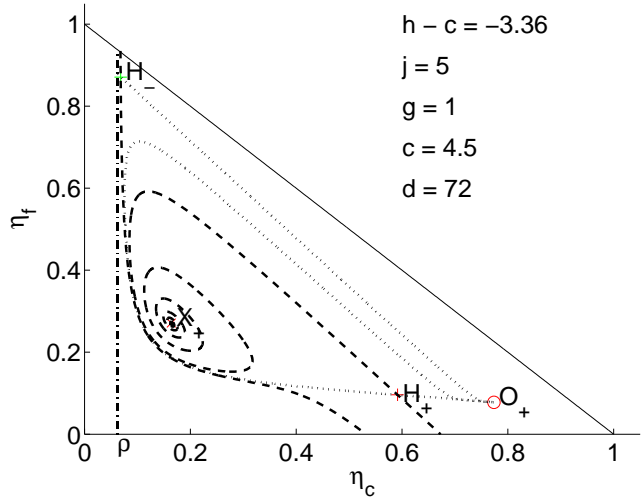}
\caption[]{Map. A separatrix bifurcation between the stable manifold of
$H_-$ and the unstable manifold of $H_+$ has introduced a first order
transition in the basin of attraction.} \label{fig.map.-3.36}
\end{figure}

\begin{figure}[!hbp]
\centering
\includegraphics[width=0.6\textwidth]{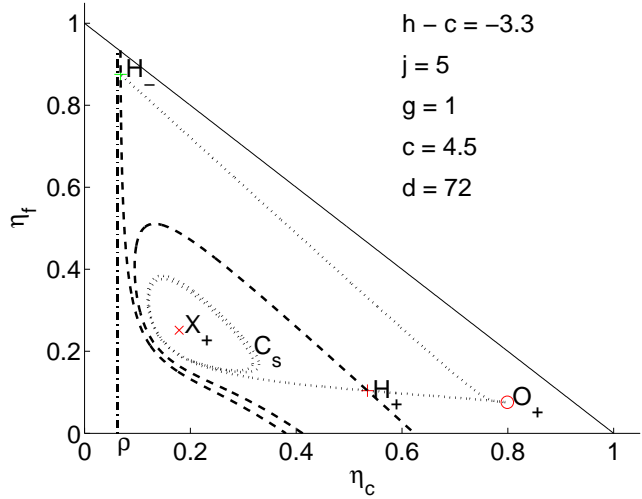}
\caption[]{Map. A separatrix bifurcation between the stable and
unstable manifolds of $H_+$ has introduced a stable cycle $C_s$ with
its own basin of attraction.} \label{fig.map.-3.3}
\end{figure}

\begin{figure}[!hbp]
\centering
\includegraphics[width=0.6\textwidth]{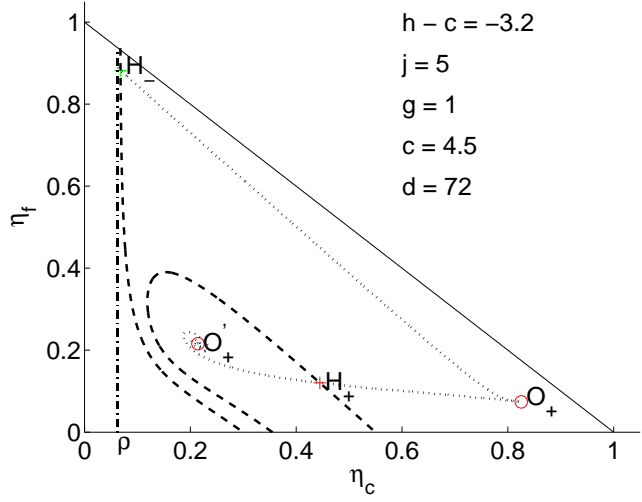}
\caption[]{Map. The stable limit cycle $C_s$ has merged with the
source $X_+$ through a Hopf bifurcation to become a sink $O_+'$.}
\label{fig.map.-3.2}
\end{figure}

\begin{figure}[!hbp]
\centering
\includegraphics[width=0.6\textwidth]{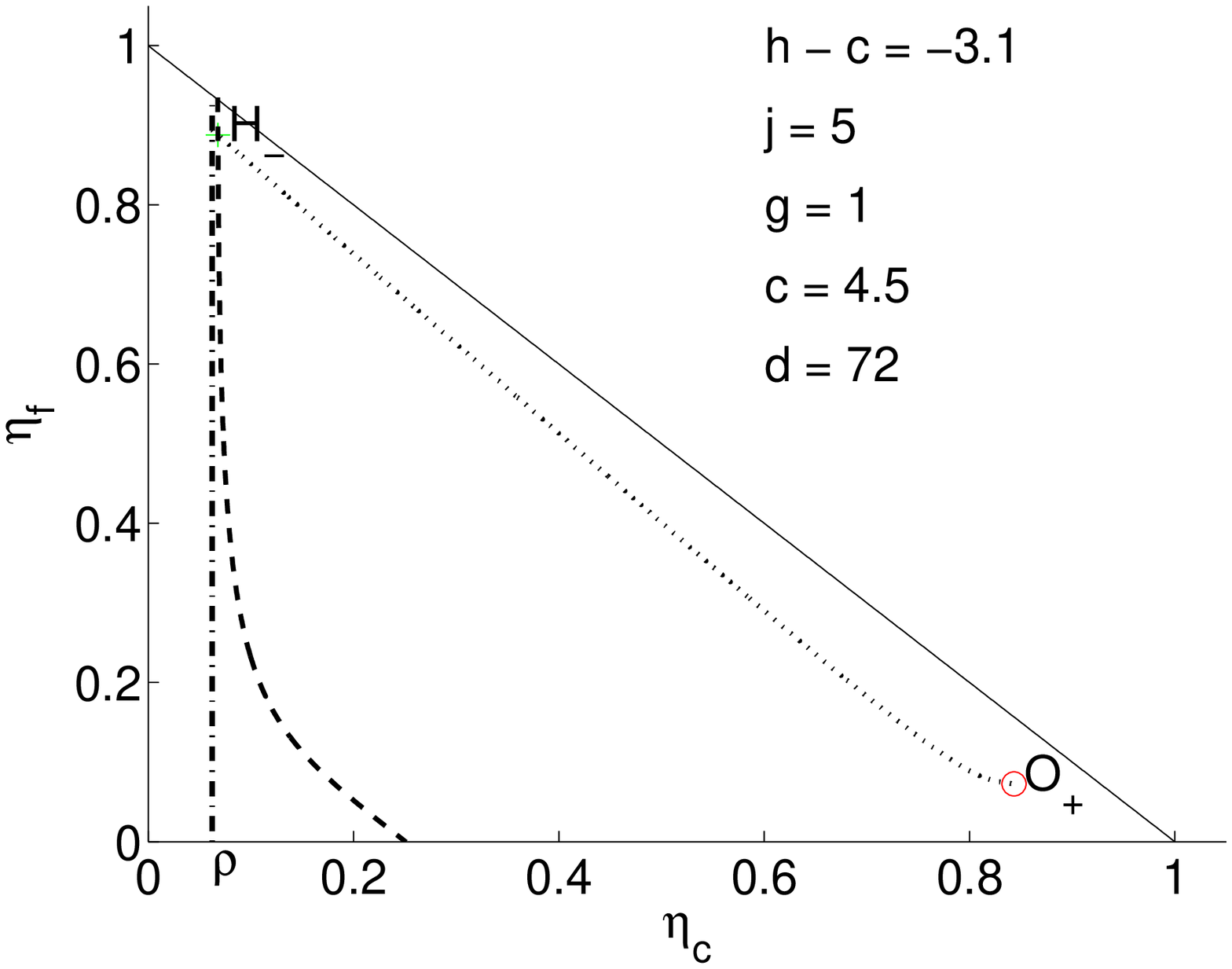}
\caption[]{Map. A final saddle node bifurcation has annihilated
$O_+'$ and $H_+$.} \label{fig.map.-3.1}
\end{figure}

In view of their crucial importance, we have numerically computed
the stable and unstable manifolds for each saddle point, with the
same set of $j$, $g$, $c$ and $d$ as in Figure
\ref{fig.GraphicSolution}, while changing $h-c$ from $-4$ to $-3$
in steps of $0.02$. The algorithms used to compute the stable and
unstable manifolds can be found in
\cite{EnglandKrauskopfOsinga2004} and Chapter 10 of
\cite{Alligood}, respectively. The results classified according to
the qualitative dynamical features are shown in
Figures~(\ref{fig.map.-3.7}-\ref{fig.map.-3.1}). Hereafter we
provide the increasing values of $h-c$, denoted by $\lambda_i$, at
which the successive bifurcations appear, together with the
qualitative descriptions of their nature.

\begin{itemize}
\item $\lambda_1=-4.0971$. For $h-c<\lambda_1$ there exists no
intersection between $\eta_1(G)$ and $\eta_2(G)$ and all the
points in $\mathbf{S}$ are eventually mapped to the trivial fixed
point, which is always a solution of the equations. At
$h-c=\lambda_1$, $\eta_2(G)$ begins to intersect $\eta_1^-(G)$ at
two points, one of which is a saddle ($H_-$) and the other is a
source ($X_-$) (Figure \ref{fig.map.-3.7}). The process of
simultaneous creation or elimination of a pair of equilibrium
points is called a saddle-node bifurcation. Note that this
bifurcation does not alter the overall behavior of the system,
since no new attractor is created. Thus, only the fixed points
$\eta_c=0$, $\eta_f > 0$ exist, and since $j>j_B$, the system may
flow to either of the two fixed points for $\eta_f$ depending on
the initial conditions.

\item $\lambda_2=-3.4201$. Another saddle-node bifurcation happens
on $\eta_1^+(G)$ at $h-c=\lambda_2$, with a saddle ($H_+$) and a
sink ($O_+$) created simultaneously (Figure \ref{fig.map.-3.4}).
Now the stable manifolds of $H_+$ and $H_-$ possess a common end
emanating from $X_+$, and together they divide $\mathbf{S}$ into
two regions. Points outside the region bounded by the stable
manifolds should converge to the trivial fixed point $\eta_c=0,
\eta_f >0$ as before, and those inside to $O_+$.

\item $\lambda_3=-3.3617$. For $\lambda_2<h-c<\lambda_3$, the
unstable manifold of $H_+$ approaches the stable manifold of $H_-$
until they coincide at $h-c=\lambda_3$ (the common manifold is
called a separatrix between $H_+$ and $H_-$). Beyond this value,
the unstable manifold of $H_+$ folds back and converges to $O_+$
while the stable manifold of $H_-$ hits the simplex boundary
$\eta_f=0$
(Figure \ref{fig.map.-3.36}).
Now the basin boundary is determined solely by the stable
manifold of $H_-$, causing the basin of attraction of $O_+$
to experience a sudden expansion.

\item $\lambda_4=-3.3504$. For $\lambda_3<h-c<\lambda_4$, the
stable and unstable manifolds of $H_+$ approach each other until
they coincide at $h-c=\lambda_4$. Beyond this value, the former
ends up at the boundary $\eta_f=0$
of $\mathbf{S}$ and the latter folds back into a stable limit cycle $C_s$ around
$X_+$ (Figure \ref{fig.map.-3.3}). At this transition $C_s$ is
introduced into the system as a new attractor with its own
basin of attraction delimited by the stable manifold of $H_+$.

\item $\lambda_5=-3.2726$. For $\lambda_4<h-c<\lambda_5$, $C_s$ shrinks and eventually merges with
$X_+$ at $h-c=\lambda_5$ to produce a sink $O_+'$ (Figure \ref{fig.map.-3.2}). The
process of transition between a sink and a source with the
simultaneous appearance or disappearance of a limit cycle is
called a Hopf bifurcation.

\item $\lambda_6=-3.1146$. For $\lambda_5<h-c<\lambda_6$, $O_+'$ and $H_+$
approach each other until they are annihilated at $h-c=\lambda_6$. We are left with
$H_-$ and $O_+$, the stable manifold of $H_-$ serving as the basin
boundary (Figure \ref{fig.map.-3.1}). This topology persists for arbitrarily larger values of
$h-c$.
\end{itemize}

Summarizing the analysis above, we can identify three types of
transition:
\begin{enumerate}
    \item saddle-node bifurcation;
    \item Hopf bifurcation;
    \item separatrix bifurcation between two saddle points, or one saddle point with itself.
\end{enumerate}
Type 1 can be directly determined from the number of intersections
between $\eta_1(G)$ and $\eta_2(G)$ (cf. figure
\ref{fig.GraphicSolution}), or from the fact that $\mathbf{J}_m$
has an eigenvalue $+1$. Due to the lack of symmetry in our system,
transcritical and pitchfork bifurcations never occur
(\cite{GuckenheimerHolmes1990} \S3.4). In case that $\mathbf{J}_m$
has an eigenvalue $-1$, one can have a period-doubling bifurcation
(\cite{GuckenheimerHolmes1990} \S3.5). This never happens in our
map, since one can easily verify that
\begin{equation}\label{eq:no_neg_eig}
    \mathbf{J}_m(1,1)>0, \;\;\; \mathbf{J}_m(2,2)>0, \;\;\; \Delta(\mathbf{J}_m)>0,
\end{equation}
which excludes the possibility that $\mathbf{J}_m$ has a negative eigenvalue.

Types 2 and 3 cannot be observed from the intersections between
$\eta_1(G)$ and $\eta_2(G)$ and must be determined numerically.
Type 2 remains a local bifurcation with $\mathbf{J}_m$ having a
pair of conjugate complex eigenvalues with unit modulus. In
contrast, Type 3 is a global bifurcation with the area of the
basin of attraction experiencing a sudden jump or, in physicist's
terms, a first-order transition. Numerically we have found no
evidence for homoclinic intersection leading to chaos, but the
question as to whether chaos exists in our two-dimensional map
remains open.

\subsection{Sequential dynamics: two-dimensional flow}
In sequential dynamics, one updates one agent chosen at random in
each time step. It is usual to consider $N$ (the number of agents
in the system) successive random sequential updates (called one
Monte-Carlo step) as being comparable to one step of parallel
updating.

In each individual update, the expected displacement of the pair
$(\eta_c,\eta_f)$ is
\begin{equation}\label{eq:single_update_displace}
    (\overline{\Delta\eta_c},\overline{\Delta\eta_f})=\frac{1}{N}(p(\eta_c,\eta_f)-\eta_c,q(\eta_c,\eta_f)-\eta_f),
\end{equation}
where $p$ and $q$ are defined in (\ref{eq:the_map}).
Taking the continuous limit $N\rightarrow\infty$,  we have the
following set of differential equations
\begin{subequations}\label{eq:pde_com}
\begin{gather}
    d\eta_c/dt=p(\eta_c,\eta_f)-\eta_c,\\
    d\eta_f/dt=q(\eta_c,\eta_f)-\eta_f,
\end{gather}
\end{subequations}
with the time unit being one Monte-Carlo step. Now the
community evolves as an autonomous system with planar phase space,
or a two-dimensional flow in $\mathbf{S}$.

The three generic types of equilibria, namely source, sink and
saddle exist in a structurally stable two-dimensional flow as
well. However, the nature of an equilibrium point
$(\eta_{c0},\eta_{f0})$ is now determined by the Jacobian matrix
of the flow
\begin{displaymath}\label{eq:jac_flow}
    \mathbf{J}_f=\left[\begin{array}{cc}
    \partial p(\eta_{c0},\eta_{f0})/\partial\eta_c-1 &
    \partial p(\eta_{c0},\eta_{f0})/\partial\eta_f\\
    \partial q(\eta_{c0},\eta_{f0})/\partial\eta_c &
    \partial q(\eta_{c0},\eta_{f0})/\partial\eta_f-1
    \end{array} \right].
\end{displaymath}
Denoting the eigenvalues of $\mathbf{J}_f$ as $a_1$ and $a_2$,
then the point is a sink if $a_1<0$ and $a_2<0$, a source if
$a_1>0$ and $a_2>0$ and a saddle if $a_1a_2<0$
(\cite{GuckenheimerHolmes1990} \S1.2$\sim$1.3). Note that the
condition for an equilibrium to be a saddle gives the same
inequality as in the parallel case. Therefore, the set of saddle
points in sequential updating coincides with that in parallel
updating. However, the set of sinks or sources are not necessarily
the same in both cases, since the governing inequalities are quite
different. The definitions of the stable and unstable manifolds
are analog to those in the two-dimensional map with the discrete
time step replaced by the continuous time variable $t$. However,
we note that the two-dimensional flow system, according to the
Poincar\'{e}-Bendixson Theorem (\cite{GuckenheimerHolmes1990}
Theorem 1.8.1), will never go into chaos.

\begin{figure}[!hbp]
\centering
\includegraphics[width=0.6\textwidth]{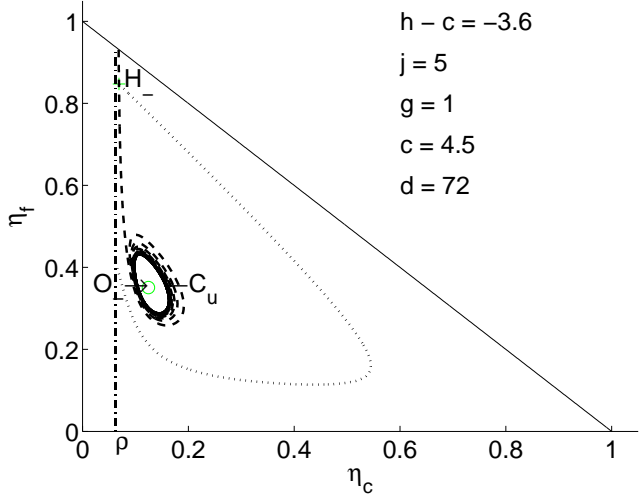}
\caption[]{Flow. A sink $O_-$ together with an unstable limit
cycle $C_u$ has emerged from the source $X_-$ through a Hopf
bifurcation.} \label{fig.flow.-3.6}
\end{figure}

\begin{figure}[!hbp]
\centering
\includegraphics[width=0.6\textwidth]{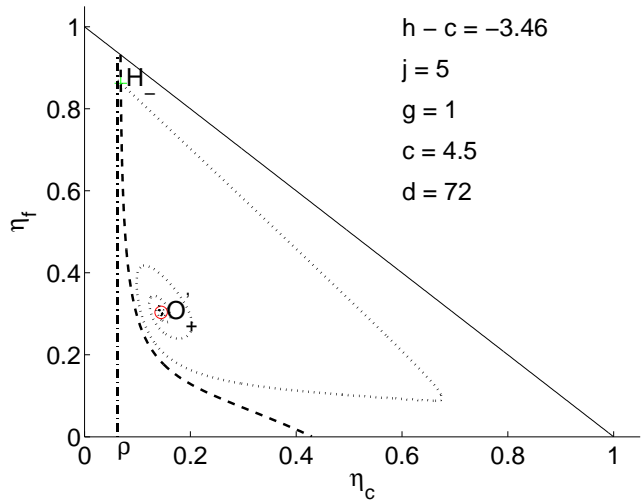}
\caption[]{Flow. A separatrix bifurcation has eliminated $C_u$ and
expanded the basin of attraction of the sink.}
\label{fig.flow.-3.46}
\end{figure}

\begin{figure}[!hbp]
\centering
\includegraphics[width=0.6\textwidth]{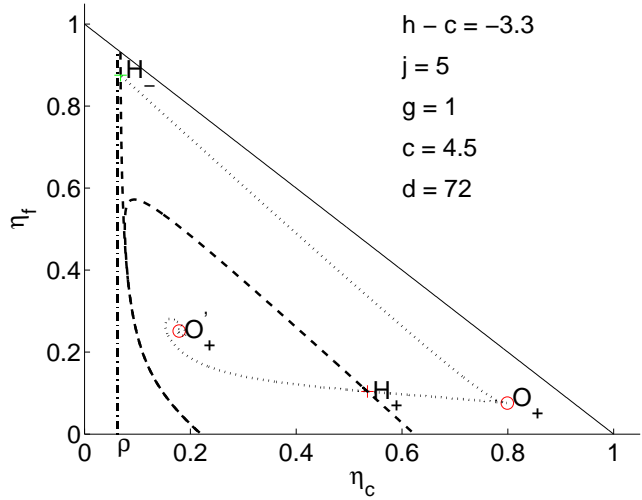}
\caption[]{Flow. $H_+$ and $O_+$ have been created from a saddle node
bifurcation on $\eta_1^+(G)$. Compare with Figure \ref{fig.map.-3.3}.} \label{fig.flow.-3.3}
\end{figure}

\begin{figure}[!hbp]
\centering
\includegraphics[width=0.6\textwidth]{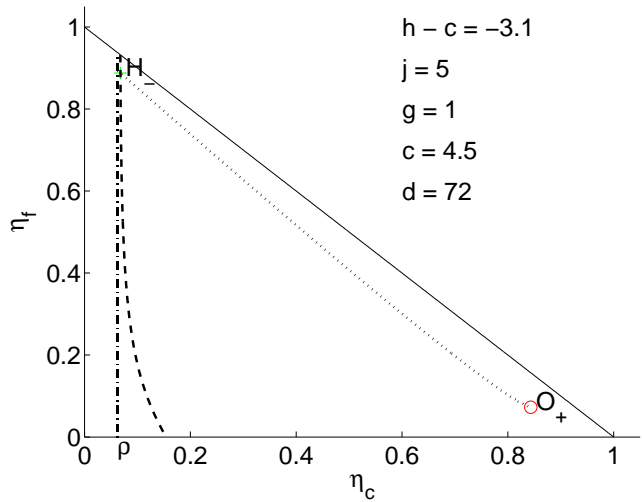}
\caption[]{Flow. A final saddle node bifurcation has annihilated $O_+'$ and
$H_+$. Compare with Figure \ref{fig.map.-3.1}.} \label{fig.flow.-3.1}
\end{figure}

We have computed the stable and unstable manifolds as for the map.
The results shown in
Figures~(\ref{fig.flow.-3.7}-\ref{fig.flow.-3.1}). Hereafter we
provide the bifurcation values of $h-c$, denoted by $\mu_i$,
together with the qualitative descriptions of the nature of the
bifurcation:

\begin{figure}[!hbp]
\centering
\includegraphics[width=0.6\textwidth]{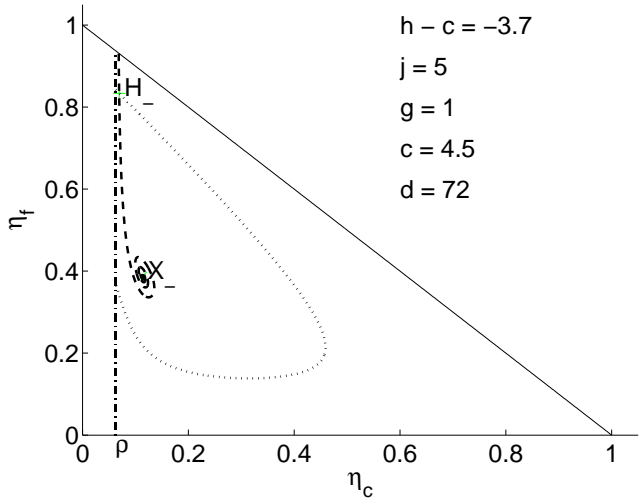}
\caption[]{Flow. The stable and unstable manifolds for the flow.
Compare with Figure \ref{fig.map.-3.7}.} \label{fig.flow.-3.7}
\end{figure}

\begin{itemize}
\item $\mu_1=\lambda_1=-4.0971$. As in the parallel case, a saddle node
bifurcation occurs at $h-c=\mu_1$ and produces a saddle $H_-$
and a source $X_-$ (Figure \ref{fig.flow.-3.7}). \item
$\mu_2=-3.6172$. A Hopf bifurcation occurs at $h-c=\mu_2$
and produces an unstable limit cycle $C_u$ together with a sink
$O_-$ from the source $X_-$ (Figure \ref{fig.flow.-3.6}). The basin of
attraction of $O_-$ is the area surrounded by $C_u$. Note
that we've been able to choose $h-c$ such that the sink remains in the negative
branch, which implies that variation of $\eta$ with respect to
$h-c$ does not always qualify to determine the stability of the fixed
point. \item $\mu_3=-3.4980$. A separatrix
bifurcation of the stable and unstable manifolds of $H_-$
eliminates $C_u$ (Figure \ref{fig.flow.-3.46}). \item
$\mu_4=\lambda_2=-3.4201$. A saddle node bifurcation occurs
at $h-c=\mu_4$ and produces a sink $O_+$ and a saddle $H^+$ (Figure
\ref{fig.flow.-3.3}). \item $\mu_5=\lambda_6=-3.1146$. A saddle node
bifurcation occurs at $h-c=\mu_5$ and annihilates the saddle $H_+$ and the source $O_+'$.
The topology persists for larger $h-c$ (Figure \ref{fig.flow.-3.1}).
\end{itemize}

\begin{figure}[!hbp]
\centering
\includegraphics[width=0.6\textwidth]{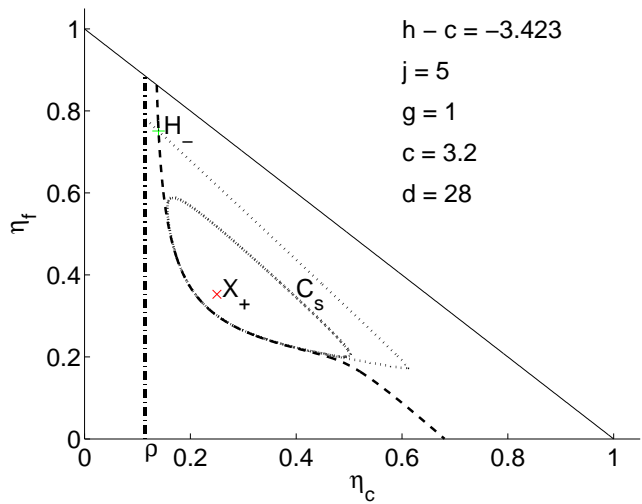}
\caption[]{Flow. A stable limit cycle in the two-dimensional flow.} \label{fig.flow.c_s}
\end{figure}

Thus, we have the same types of bifurcation as in the parallel
case, and Peixoto's Theorem (\cite{GuckenheimerHolmes1990} Theorem
1.9.1) indicates that these are the only types of bifurcation
expected in a two-dimensional flow system. However, let's note
that the overall topology of the dynamics depends strongly on
their order of occurrence. For example, if the separatrix
bifurcation happens before the Hopf bifurcation, a stable limit
cycle can be created just like in the parallel case (Figure
\ref{fig.flow.c_s}).

\section{Comparison with Computer Simulations}
\label{sec:computer simulations}
\begin{figure}[!hbp]
\centering
\includegraphics[width=0.6\textwidth]{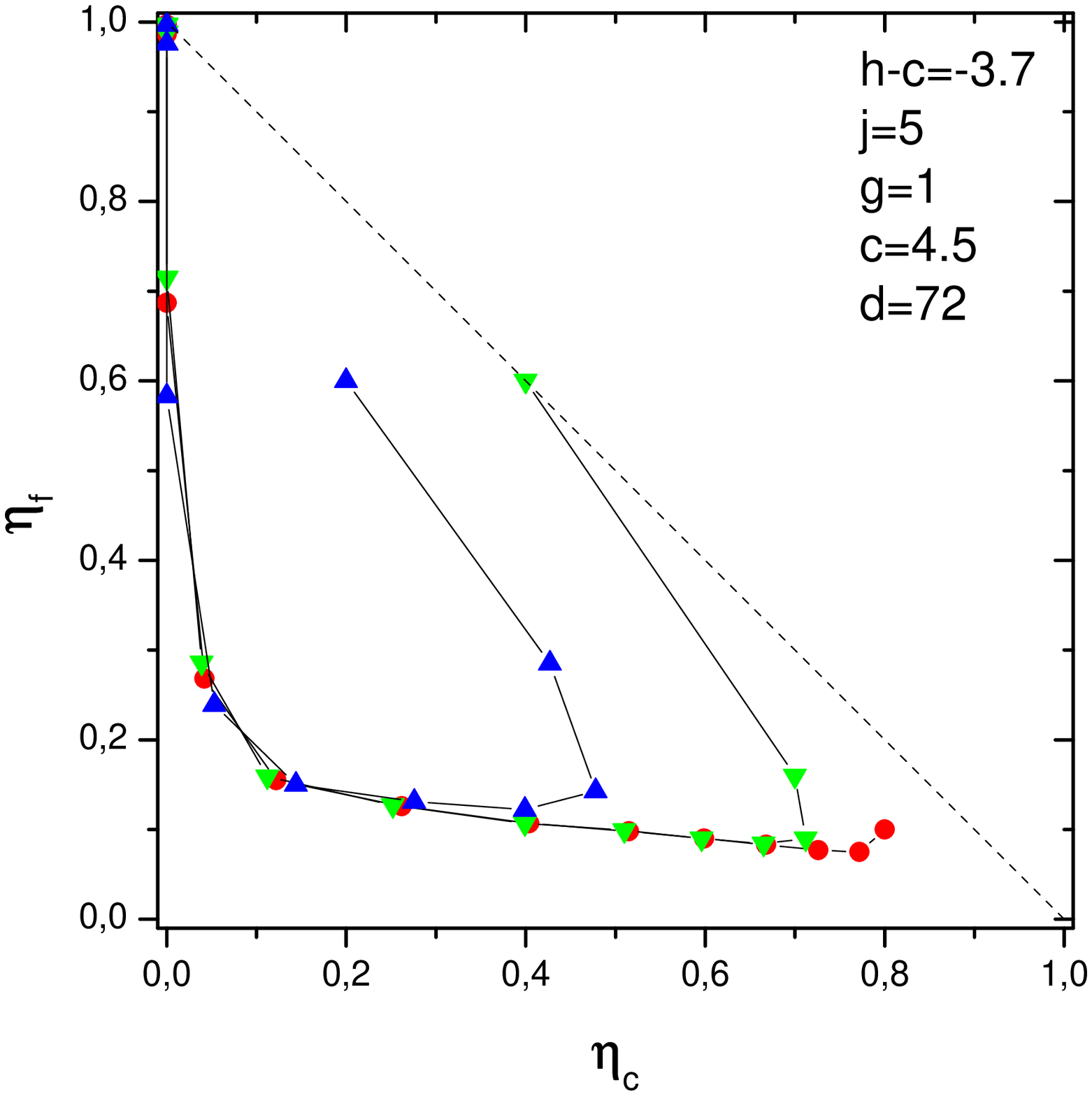}
\caption[]{Examples of trajectories of a simulated system, starting from different initial conditions, corresponding to the map of figure \protect{\ref{fig.map.-3.7}}.}
\label{fig.sim_map.-3.7}
\end{figure}

\begin{figure}[!hbp]
\centering
\includegraphics[width=0.6\textwidth]{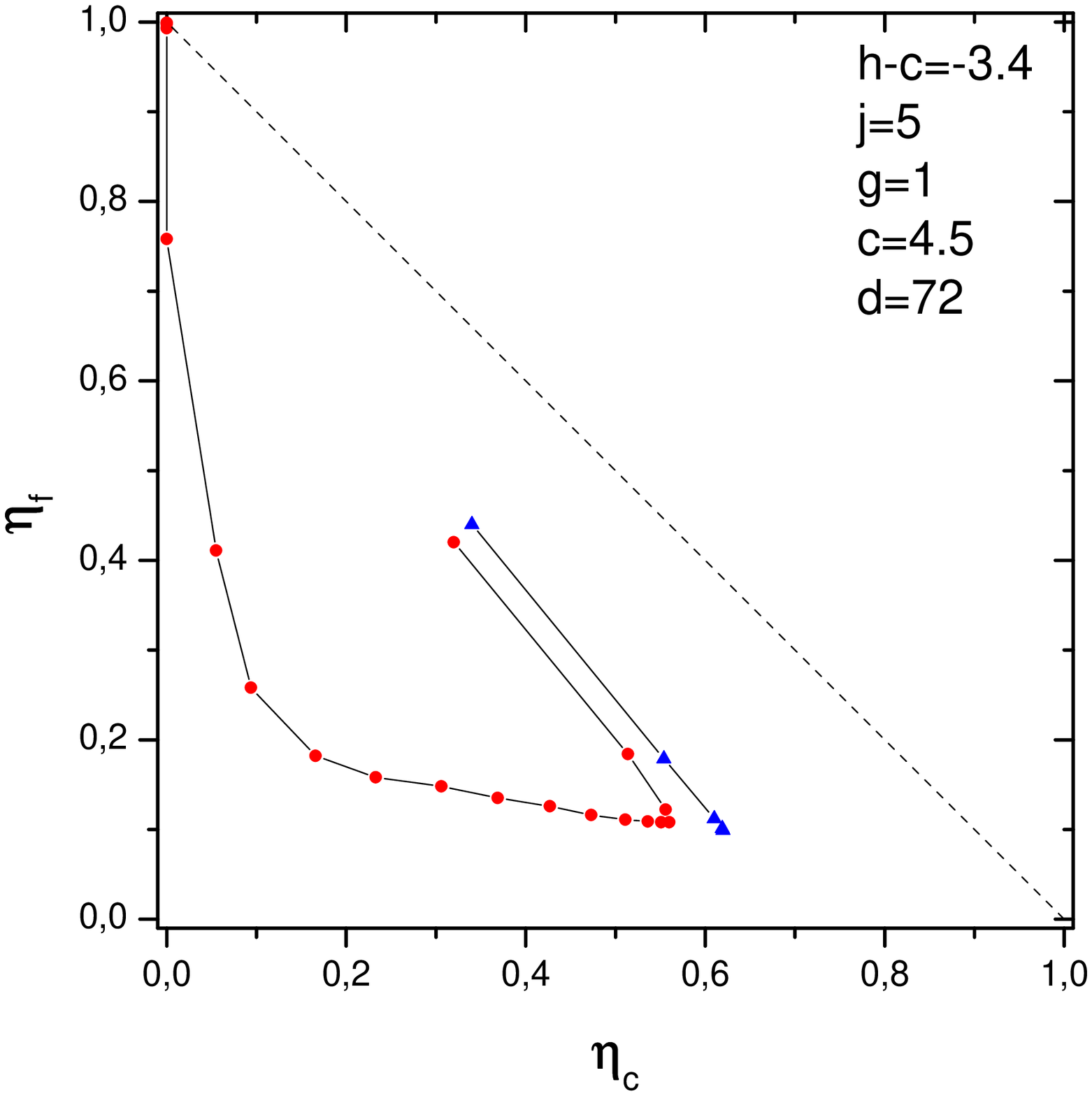}
\caption[]{Examples of trajectories of a simulated system, starting from different initial conditions, corresponding to the map of figure \protect{\ref{fig.map.-3.4}}.}
\label{fig.sim_map.-3.4}
\end{figure}

\begin{figure}[!hbp]
\centering
\includegraphics[width=0.6\textwidth]{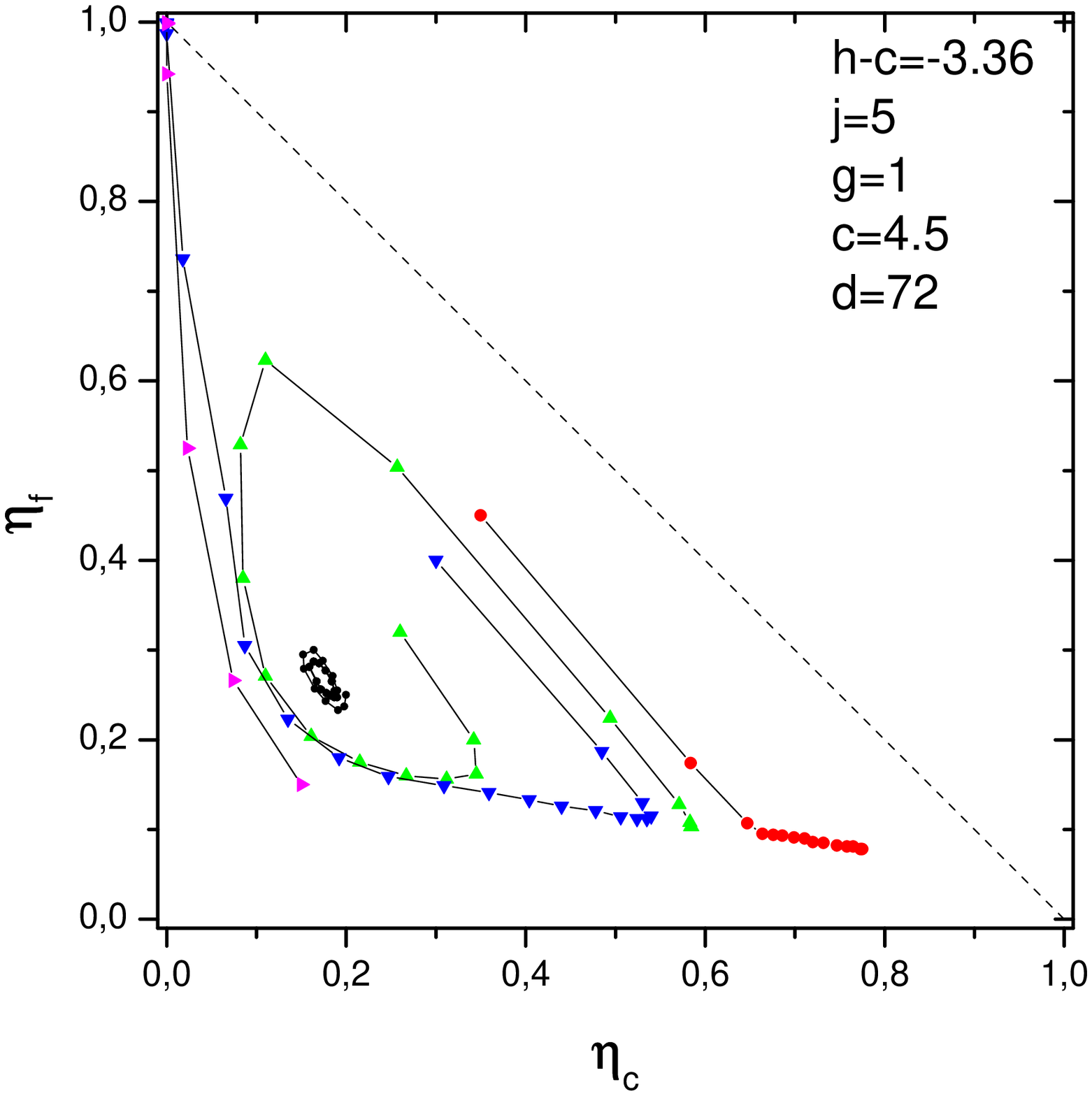}
\caption[]{Examples of trajectories of a simulated system, starting from different initial conditions, corresponding to the map of figure \protect{\ref{fig.map.-3.36}}.}
\label{fig.sim_map.-3.36}
\end{figure}

\begin{figure}[!hbp]
\centering
\includegraphics[width=0.6\textwidth]{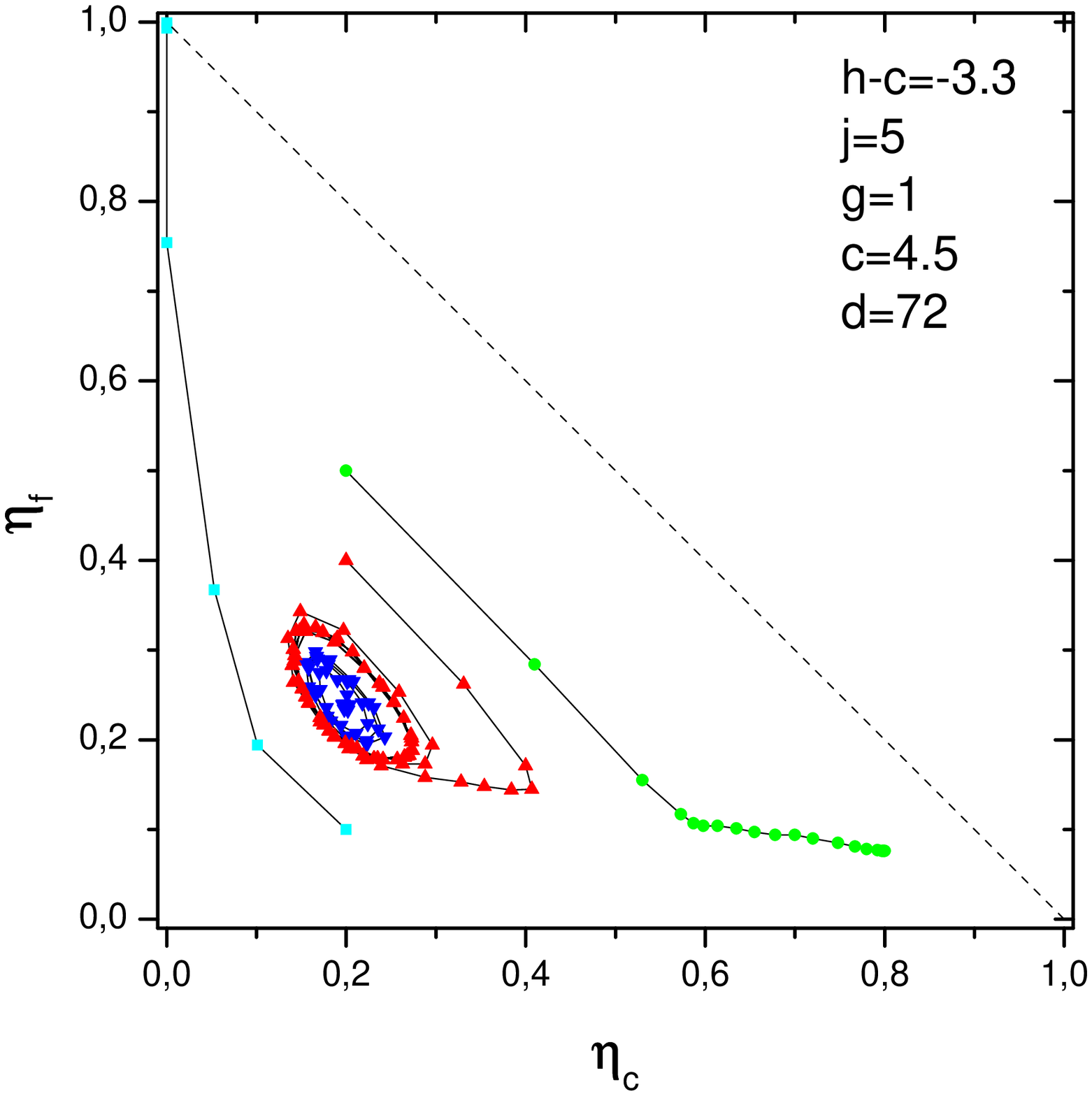}
\caption[]{Examples of trajectories of a simulated system, starting from different initial conditions, corresponding to the map of figure \protect{\ref{fig.map.-3.3}}.}
\label{fig.sim_map.-3.3}
\end{figure}

\begin{figure}[!hbp]
\centering
\includegraphics[width=0.6\textwidth]{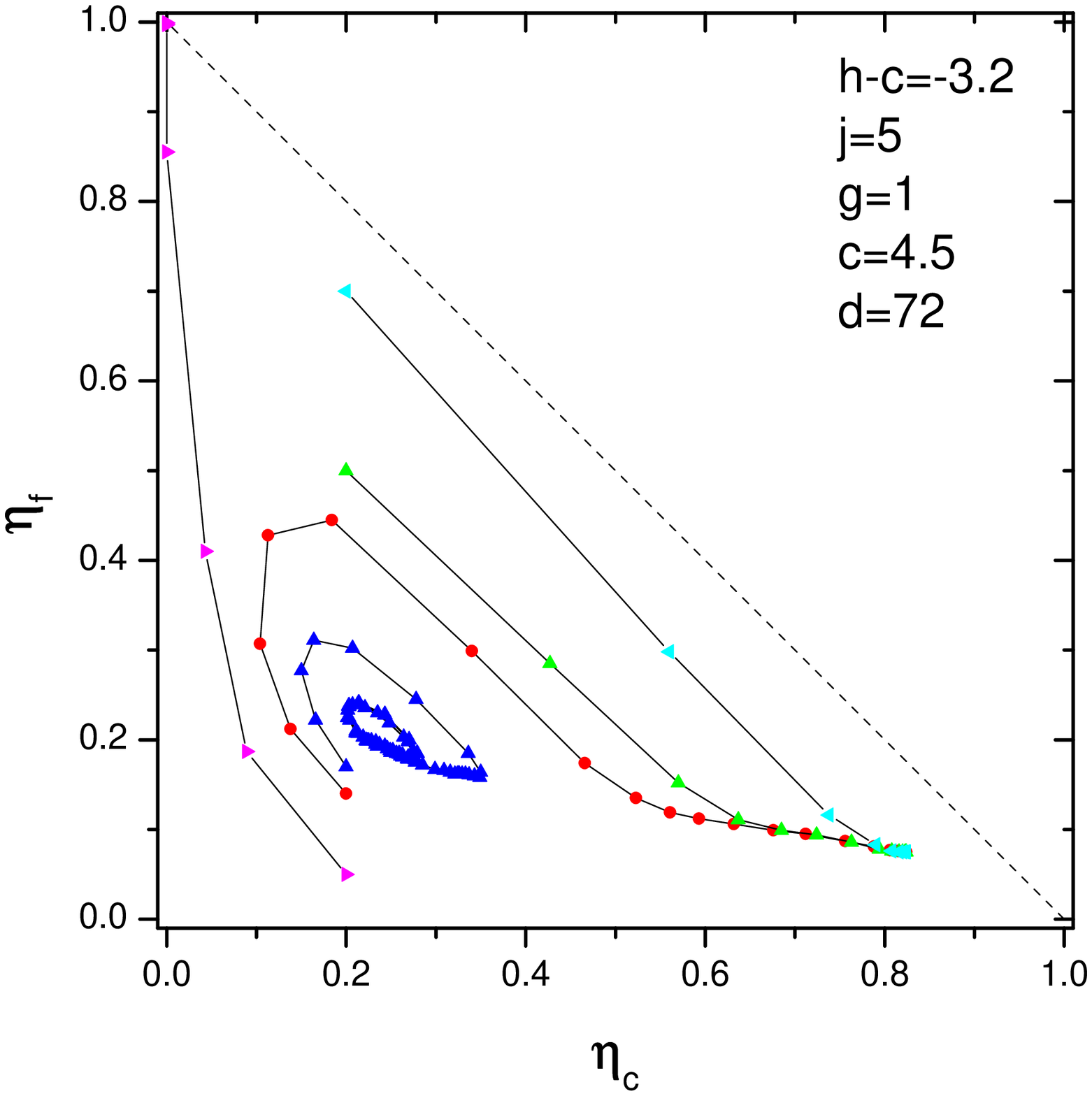}
\caption[]{Examples of trajectories of a simulated system, starting from different initial conditions, corresponding to the map of figure \protect{\ref{fig.map.-3.2}}.}
\label{fig.sim_map.-3.2}
\end{figure}

\begin{figure}[!hbp]
\centering
\includegraphics[width=0.6\textwidth]{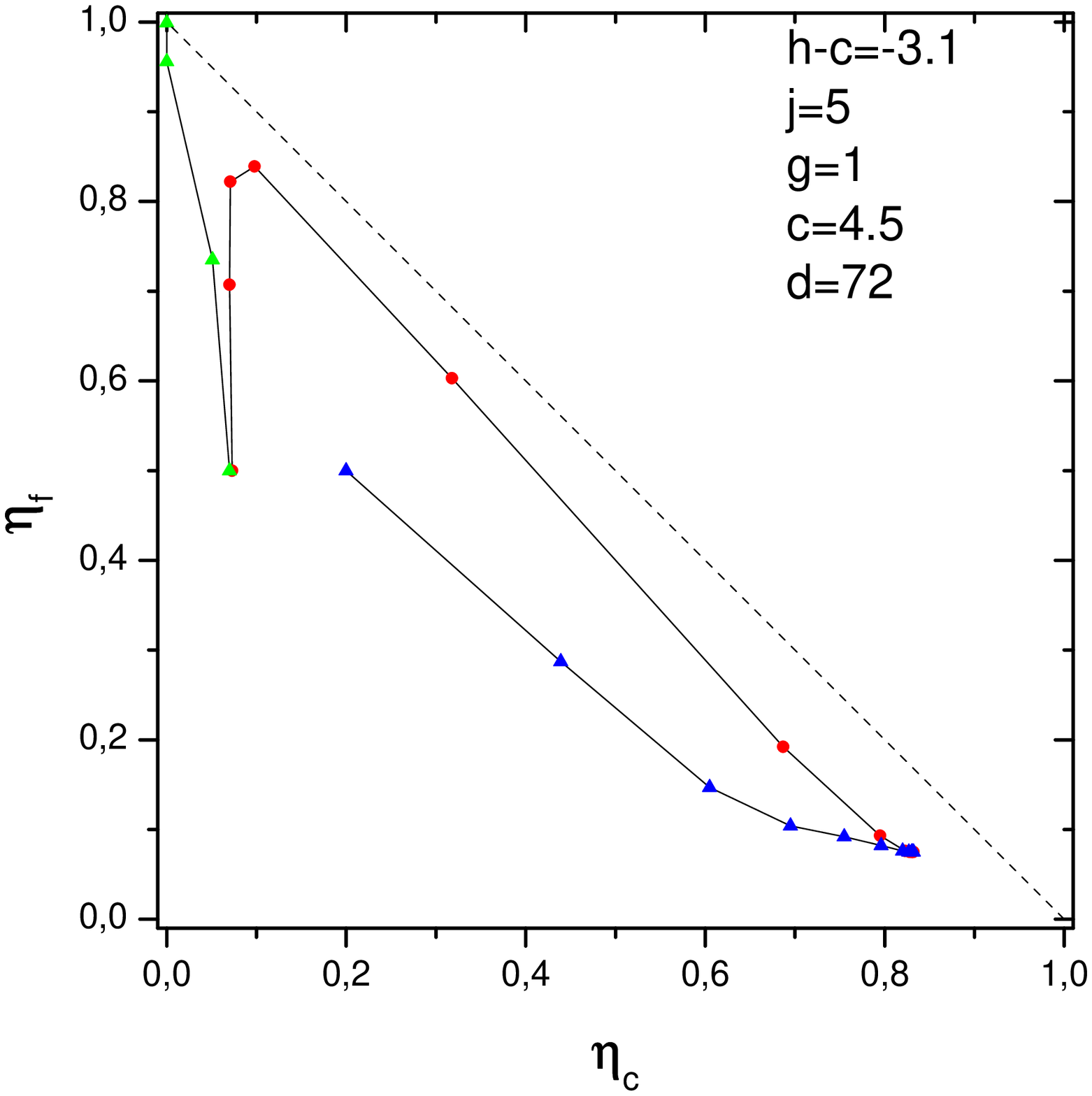}
\caption[]{Examples of trajectories of a simulated system, starting from different initial conditions, corresponding to the map of figure \protect{\ref{fig.map.-3.1}}.}
\label{fig.sim_map.-3.1}
\end{figure}

We simulated systems of $N=1000$ agents, corresponding to the parameter values of figures \ref{fig.sim_map.-3.7} to \ref{fig.sim_map.-3.1}, with parallel dynamics. Starting from different initial conditions, we determined the trajectories in the phase space $(\eta_c,\eta_f)$. The results, presented on figures \ref{fig.sim_map.-3.7} to \ref{fig.sim_map.-3.1} show an excellent egreement with the analytical results. The complexity of the map diagram is reflected on the winding trajectories. One of the most striking results of the simulations is that the fraction of cooperators and free-riders oscillate in time, even in systems for which do not expect to have cycles.

\section{Conclusion}
In this paper we have studied the time evolution
of a social system with cooperators and free-riders.
We analyzed the two-dimensional dynamical system
and obtained different phase diagrams corresponding
to various parameter settings. The fact that a rich
class of bifurcations can occur within the narrow
tubular region enclosed by the curves $h-c=-3.7$
and $h-c=-3.1$ in Figure (\ref{fig.GraphicSolution})
is in itself truly remarkable. Physically this
phenomenon has its root in the strong asymmetry
inherent in the interaction between agents, due to
the idiosyncratic weights that the free-riders give
to the social disapproval. Moreover, in the context
of parallel and sequential updating, we have sometimes
distinctly different phase diagram for the same
parameter setting. A natural question to ask is what
happens in the intermediate case, i.e. some agents'
decisions are taken simultaneously while
others' decisions are taken independently of each
other.
Numerical simulations show that the analytical results
obtained in the thermodynamic limit are valid for systems
of $N=1000$ individuals. The dynamical behaviour of the
simulated systems show winding trajectories in very large
regions of the phase diagram. Oscillations of the fraction
of cooperators and free-riders are thus expected in such
systems. We are currently performing simulations with
sequential dynamics. Already obtained results also
exhibit the above mentionned oscillatory behaviour.
Future work can also focus on the dynamics
in the case of non-global neighborhood, which
probably requires more sophisticated techniques
from the theory of dynamical systems.

\bibliographystyle{unsrt}

\begin{thebibliography}{777}

\addcontentsline{toc}{section}{References}

\bibitem{PhWaGoNa05} Denis Phan, Roger Waldeck, Mirta B. Gordon and Jean-Pierre Nadal (2005),
{\it Adoption and cooperation in communities: mixed equilibrium in
polymorphic populations}; Annual Workshop on Economics with
Heterogeneous Interacting Agents - WEHIA 2005, June 13-15,
University of Essex, UK
\bibitem{GoPhWaNa05} M. B. Gordon, D. Phan, R. Waldeck and J.-P. Nadal (2005),
{\it Cooperation and free-riding with moral cost}; Proceedings of
International Conference on Cognitive Economics (ICCE),
Sofia-Bulgaria
\bibitem{FehrGachter2002} E. Fehr and S. Gaechter (2002),
{\it Altruistic Punishment in Humans}; Nature 415 137-140
\bibitem{GaechterFehr1999} S. Gaechter and E. Fehr (1999),
{\it Collective Action as a Social Exchange}; Journal of Economic
Behavior and Organization 39 341-369
\bibitem{Durlauf01} S. N. Durlauf (2001),
{\it A framework for the study of individual behaviour and social
interactions}; Working paper
\bibitem{GoNaPhVa05} M. B. Gordon, J.-P. Nadal, D. Phan and J.
Vannimeuns (2005), {\it Seller's dilemma due to social
interactions between customers}; Physica A 356, Issues 2-4 628-640
\bibitem{NaPhGoVa05} J-P. Nadal, D. Phan, M. B. Gordon and J.
Vannimenus (2005), {\it Multiple equilibria in a monopoly market
with heterogeneous agents and externalities}; Quantitative
Finance, to be published
\bibitem{GuckenheimerHolmes1990} J. Guckenheimer and P. Holmes (1990),
{\it Nonlinear oscillations, dynamical systems, and bifurcations
of vector fields}; Springer-Verlag
\bibitem{EnglandKrauskopfOsinga2004} J. P. England, B. Krauskopf and H. M.
Osinga (2004), {\it Computing One-Dimensional Stable Manifolds and
Stable Sets of Planar Maps without the Inverse}; SIAM Journal on
Applied Dynamical Systems
\bibitem{Alligood} K. Alligood, T. Sauer and J.A. Yorke (1997),
{\it Chaos: An Introduction to Dynamical Systems}; Springer-Verlag

\end{thebibliography}

\end{document}